\documentclass{PoS}

\usepackage{cite}

\title{Optimizing the green-field beta beam: \\ Small versus large $\theta_{13}$}

\ShortTitle{Optimizing the green-field beta beam}

\author{\speaker{Walter Winter}\thanks{Supported by the Emmy Noether program of Deutsche Forschungsgemeinschaft (DFG).}
\\
        Institut f{\"u}r theoretische Physik und Astrophysik, Universit{\"a}t W{\"u}rzburg, D-97074 W{\"u}rzburg \\
        E-mail: \email{winter@physik.uni-wuerzburg.de}}

\abstract{We discuss the optimization of a green-field beta beam in terms of baseline, boost factor, luminosity, and isotope pair used. We identify two qualitatively different cases: $\theta_{13}$ not discovered at the time a decision has to be made ($\theta_{13}$ small), and $\theta_{13}$ discovered at that time ($\theta_{13}$ large). For small $\theta_{13}$, it turns out that the obtainable sensitivity is essentially a matter of the effort one is willing to spend. For large $\theta_{13}$, however, one can find clear optimization criteria, and one can use the information on $\theta_{13}$ obtained until then.
}

\FullConference{10th International Workshop on Neutrino Factories, Super beams and Beta beams \\
		 June 30 - July 5 2008 \\
		 Valencia, Spain}

\newcommand{\ie}{{\it i.e.}}

\newcommand{\etc}{{\it etc.}}
\newcommand{\eq}{Eq.}

\newcommand{\fig}{Fig.}

\newcommand{\Ref}{Ref.}
\newcommand{\Refs}{Refs.}

\newcommand{\stheta}{\sin^22\theta_{13}}
\newcommand{\deltacp}{\delta_\mathrm{CP}}

\newcommand{\equ}[1]{\eq~(\ref{equ:#1})}
\newcommand{\figu}[1]{\fig~\ref{fig:#1}}

\newcommand{\bi}{\begin{itemize}}
\newcommand{\ei}{\end{itemize}}

\newcommand{\brli}{($^8$B,$^8$Li)}
\newcommand{\nehe}{($^{18}$Ne, $^6$He)}

\begin{document}

%\section{Introduction - green field scenario}

Beta beams~\cite{Zucchelli:2002sa,Mezzetto:2003ub,Autin:2002ms,Bouchez:2003fy,Lindroos:2003kp} produce a neutrino beam by the decay of radioactive isotopes in straight sections of a storage ring. They have been studied in specific scenarios from low to very high $\gamma$'s~\cite{BurguetCastell:2003vv,BurguetCastell:2005pa,Agarwalla:2005we,Campagne:2006yx,Donini:2006dx,Donini:2006tt,Agarwalla:2006vf,Agarwalla:2007ai,Coloma:2007nn,Jansson:2007nm,Meloni:2008it,Agarwalla:2008ti}. In this talk, we discuss the green-field optimization of a beta beam, as it has been performed in \Refs~\cite{Huber:2005jk,Agarwalla:2008gf,Winter:2008cn}. Hereby, ``green-field scenario'' means that no specific accelerator, baseline $L$, boost factor $\gamma$, or isotope pair \nehe\ or \brli\ is assumed. We will typically assume $1.1 \cdot 10^{18}$ useful ion decays/year for neutrinos and $2.9 \cdot 10^{18}$ useful ion decays/year for anti-neutrinos, where the experiment is operated five years in the neutrino mode and five years in the antineutrino mode. In addition, we use a $500 \, \mathrm{kt}$ (fiducial mass) water Cherenkov detector or a $50 \, \mathrm{kt}$ (fiducial mass) magnetized iron calorimeter. These standard numbers will be referred to as a {\bf luminosity scaling factor} $\mathcal{L}=1$, which depends on the detector technology used. Note that $\mathcal{L}$ scales the number of useful ion decays/year $\times$ running time $\times$ detector mass $\times$ detector efficiency. The goal will be to optimize the free parameters (such as isotope pair, luminosity, $L$, and $\gamma$) for the best physics potential. Note that we only discuss two specific detector technologies for the sake of simplicity here.

For a qualitative discussion of the beta beam spectrum, note that the peak energy is approximately given by $\gamma \cdot E_0$ and the maximum energy is approximately given by $2 \cdot \gamma \cdot E_0$, where $E_0$ is the endpoint energy of the decay. The total flux, on the other hand, is approximately proportional to $N_\beta \cdot \gamma^2$, where $N_\beta$ is the number of useful ion decays. Comparing different isotope pairs with different endpoint energies, one can relate these to each other by postulating a similar spectrum, leading to the same cross sections, baseline, physics (such as the MSW effect), \etc. Obviously, one can either use isotopes with lower endpoint energy and a higher $\gamma$, or vice versa. If one in addition requires a similar total flux, one obtains from the above relations that
\begin{equation}
\frac{N_\beta^{(1)}}{N_\beta^{(2)}} \simeq \left( \frac{E_0^{(1)}}
{E_0^{(2)}}\right)^2 \, , \quad 
\frac{\gamma^{(1)}}{\gamma^{(2)}} \simeq \frac{E_0^{(2)}}{E_0^{(1)}} \, ,
\label{equ:cond}
\end{equation}
where $1$ and $2$ refer to the different isotope pairs. Since $E_0$ for \brli\ is about a factor of $3.5$ higher (in average) than that of \nehe , we have 
\begin{equation} 
N_\beta^{\mathrm{(^8B,^8Li)}} \simeq 12 \cdot
N_\beta^{\mathrm{(^{18}Ne, ^6He)}} \, , \quad
\gamma^{\mathrm{(^{18}Ne, ^6He)}} \simeq 3.5 \cdot \gamma^{\mathrm{(^8B,^8Li)}}
\label{equ:condnum}
\end{equation} in order to have a similar physics output. Note that $N_\beta$ is (primarily) a source degree of freedom, whereas $\gamma$ represent the acceleration effort, it is not clear which of these two conditions dominate, and which isotope pair will be preferred in a green-field setup.

%\section{Beta beams for small $\theta_{13}$}

Let us first of all discuss beta beams for small $\theta_{13}$, where we refer to ``small $\theta_{13}$'' as values of $\theta_{13}$ not yet discovered by the reactor experiments and first generation superbeams. In this case, 
we optimize in the $\theta_{13}$ direction, which means that we require sensitivity to $\theta_{13}$, the mass hierarchy (MH), and CP violation (CPV) for as small as possible $\theta_{13}$. There are, however, two unknowns in this optimization. First of all, it is unclear for which values of (true) $\deltacp$ such an optimization should be performed. And second, how small $\theta_{13}$ is actually good enough? It turns out that, to a first approximation, the higher the $\gamma$, the better~\cite{Huber:2005jk}, unless the detector technology runs into its limitations. 
\begin{figure}
\begin{center}
\includegraphics[width=0.9\textwidth]{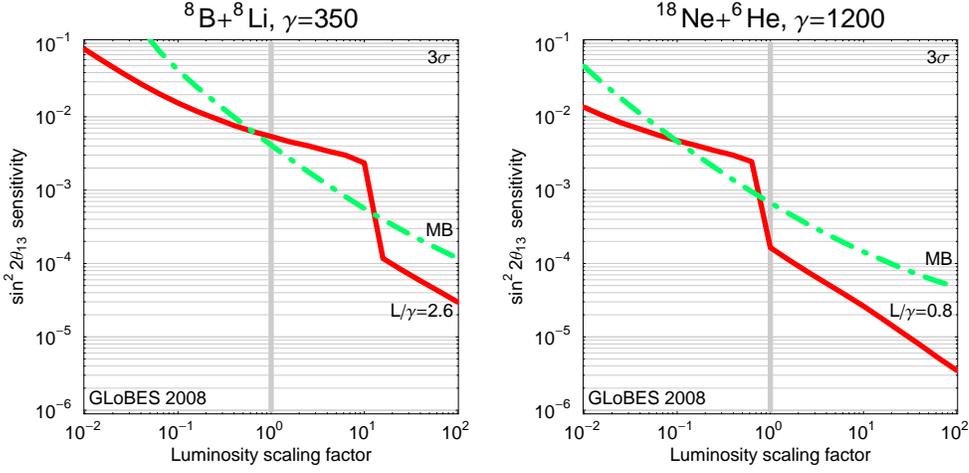}
\vspace*{-0.5cm}
\end{center}
\caption{\label{fig:theta13} The $\stheta$ sensitivity ($3 \sigma$) as a function of a
luminosity scaling factor (see main text) for a $50 \, \mathrm{kt}$ iron calorimeter.  The panels
represent the different isotopes and different $\gamma$ as
indicated in the captions. The green dashed-dotted curves
correspond to the magic baseline ``MB'' with $L=7500 \,
\mathrm{km}$ fixed, the red solid curves to a short baseline with
an $L/\gamma$ depending on the isotope. A true normal hierarchy is assumed. Figure from 
\Ref~\cite{Agarwalla:2008gf}.}
\end{figure}
In addition, the higher the luminosity, the better, as it is illustrated in \figu{theta13} for the $\stheta$ sensitivity for two different isotope pairs/$\gamma$'s, and two different baseline choices~\cite{Agarwalla:2008gf}. Therefore, the minimal reachable $\theta_{13}$ is more or less a matter of cost, and it is not possible to clearly identify a minimal setup measuring the unknown quantities.

The optimal baseline depends for any specific scenario (specific luminosity, isotope pair, and $\gamma$) on the performance indicator. For example, CP violation in general prefers shorter baselines, whereas the mass hierarchy requires strong matter effects and therefore long baselines~\cite{Huber:2005jk}. For the higher $\gamma$ options and, for instance, a iron calorimeter, two sets of suboptimal baselines can be identified~\cite{Agarwalla:2008gf}: A ``short'' baseline with $L/\gamma \simeq 0.8$ for \nehe\ or $L/\gamma=2.6$ for \brli , and the ``magic'' baseline $L \simeq 7 \, 500 \, \mathrm{km}$~\cite{Huber:2003ak} to resolve correlations and degeneracies. With this detector, in principle, the MH is best measured with a \brli\ beam at the magic baseline, whereas CPV is best measured with a \nehe\ beam at the short baseline. For the $\stheta$ sensitivity and \brli , it turns out that the magic baseline performs better for $\gamma \gtrsim 350$, whereas below that value the shorter baseline performs better (for $\mathcal{L}=1$). For the \nehe\ beam, one would prefer the short baseline in most of the cases. Note, however, that the baseline choice depends on statistics as well, as illustrated in \figu{theta13} for two different isotope pairs and $\gamma$'s. If the luminosity is different from the nominal luminosity $\mathcal{L}=1$, the optimal baseline for $\theta_{13}$ indeed changes. The kink in these scalings comes from the resolution of degeneracies with a certain threshold statistics, whereas for the magic baseline, there are no such degeneracies a priori. One can also read off from \figu{theta13} that  \equ{condnum} is satisfied: In this figure, the $\gamma$ is increased by a factor of about 3.5 from the left to the right panel, where \nehe\ instead of \brli\ is used. Indeed, one can read off from the kink at the short baseline, that for \nehe\ about a factor of ten lower luminosity is required than for \brli . Note that the $L/\gamma$ for the shorter baselines are just related by the endpoint energy ratio.

%\section{Beta beams for large $\theta_{13}$}

\begin{figure}
\begin{center}
\includegraphics[width=0.9\textwidth]{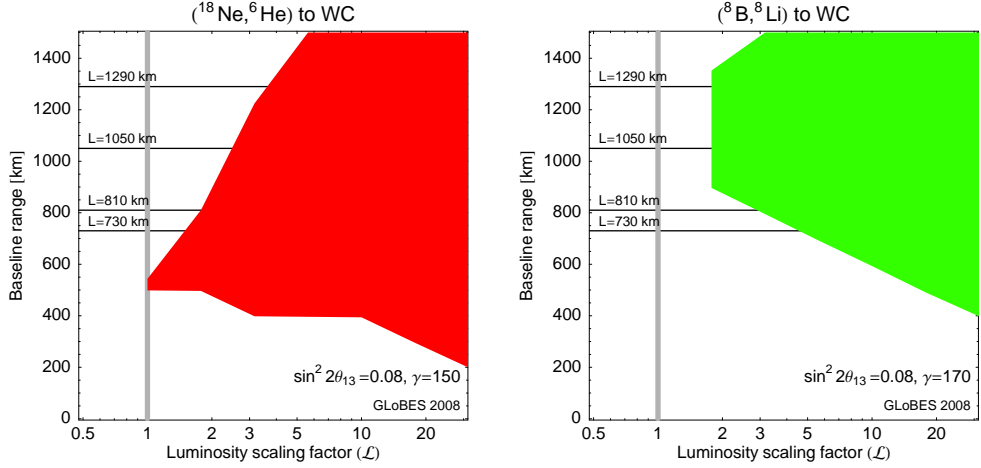}
\vspace*{-0.5cm}
\end{center}
\caption{\label{fig:ll} Possible baseline range for a \nehe , left, or \brli , right, beta beam as a function of the luminosity scaling factor $\mathcal{L}$ for a $500 \, \mathrm{kt}$ water Cherenkov detector. In these figures, $\gamma$ is fixed to 150 (left) and 170 (right), respectively. The baseline ranges are given for a Double Chooz best-fit $\sin^2 2 \theta_{13}=0.08$. The ``sensitivity'' for large $\theta_{13}$ is defined in the main text. }
\end{figure}

Compared to the small $\theta_{13}$ case, in which one optimizes for $\theta_{13}$ reaches as good as possible, the minimum wish list for small $\theta_{13}$ from the physics point of view could be rather straightforward:
A $5\sigma$ independent confirmation of $\stheta>0$,
a $3\sigma$ determination of the MH for {\em any} (true) $\deltacp$,
and a $3\sigma$ establishment of CPV for 80\% of all (true) $\deltacp$.
Since we have assumed that $\stheta$ has been measured, one can use this knowledge to
optimize the experiment. Therefore, we postulate these sensitivities 
in the {\em entire remaining allowed $\theta_{13}$ range}, which means the range remaining 
after a $\stheta$ discovery (in fact, we assume the range after three years of Double Chooz operation~\cite{Huber:2006vr}). 
In this case, one can approach the optimization of the experiment
from different points of view. For example, in \Ref~\cite{Winter:2008cn}, an optimization in the $L$-$\gamma$ plane
was performed to identify the minimal $\gamma$ for which the above performance indicators can be
measured. It has turned out that a $\gamma$ as high as 350 might not be necessary~\cite{BurguetCastell:2005pa}.
The MH sensitivity typically imposes a lower bound on the baseline $L \gtrsim 500 \, \mathrm{km}$.
The CPV sensitivity typically (for not too large luminosities) imposes a lower bound on $\gamma$. 
Compared to \Ref~\cite{Winter:2008cn}, one can also perform the optimization for a fixed $\gamma$.
For instance, we show in \figu{ll} the possible baseline range for a \nehe\ beam (left panel) and a \brli\ beam (right panel) to a $500 \, \mathrm{kt}$ water Cherenkov detector for a fixed $\gamma=150$ (left panel) and a fixed $\gamma=170$ 
(right panel), respectively, as a function of the luminosity scaling factor $\mathcal{L}$. These fixed $\gamma$'s correspond to the maximum which might be possible at the CERN SPS. As one can read off from this figure, $\mathcal{L}=1$ may not be sufficient for the \nehe\ beam, especially since sensitivity is only given in a very small baseline window.
However, if a \brli\ beam was used with a slightly more (about a factor of two) better luminosity, which may, for instance, be achieved by using a production ring for the ion production, the required sensitivities might be achievable in a relatively wide baseline range $850 \, \mathrm{km} \lesssim L \lesssim 1 \, 350 \, \mathrm{km}$. 

%\section{Summary and conclusions}

In summary, we have discussed the optimization of a green-field beta beam in terms of baseline, $\gamma$, luminosity, and isotopes used. If $\theta_{13}$ is not discovered at the time a decision for an experiment has to be made, the optimization might be primarily driven by $\stheta$ reaches as good as possible. In this case, there are no obvious criteria, such as a specific value of $\stheta$ which may be interesting, which means that the sensitivity is essentially a matter of how much effort one is willing to spend. For large $\theta$, \ie, if $\theta_{13}$ has been discovered, however, relatively objective criteria for the optimization can be found, and the knowledge on $\theta_{13}$ can be used. In this case, a beta beam with a $\gamma$ reachable by the CERN SPS could be sufficient if \brli\ with a sufficiently high luminosity was used.

%\bibliographystyle{apsrev}
%\bibliography{references}

\end{document}